\documentclass{article}
\usepackage[letterpaper]{geometry}
\usepackage{spconf,graphics,hyperref}
\usepackage{graphicx}
\usepackage{color}
\usepackage{multirow}

\title{Holoscopic 3D Micro-Gesture Database for Wearable Device Interaction}

\name{Yi Liu, Hongying Meng, Mohammad Rafiq Swash, Yona Falinie A. Gaus, Rui Qin}
\address{
Department of Electronic and Computer Engineering\\
Brunel University London, UK\\
Hongying.Meng@brunel.ac.uk}

\begin{document}
\maketitle

\begin{abstract}
With the rapid development of augmented reality (AR) and virtual reality (VR) technology, human-computer interaction (HCI) has been greatly improved for gaming interaction of AR and VR control. The finger micro-gesture is one of the important interactive methods for HCI applications such as in the Google Soli and Microsoft Kinect projects. However, the progress in this research is slow due to the lack of high quality public available database. In this paper, holoscopic 3D camera is used to capture high quality micro-gesture images and a new unique holoscopic 3D micro-gesture (HoMG) database is produced. The principle of the holoscopic 3D camera is based on the fly’s viewing system to see the objects. HoMG database recorded the image sequence of 3 conventional gestures from 40 participants under different settings and conditions. For the purpose of micro-gesture recognition, HoMG has a video subset with 960 videos and a still image subset with 30635 images. Initial micro-gesture recognition on both subsets has been conducted using traditional 2D image and video features and popular classifiers and some encouraging performance has been achieved. The database will be available for the research communities and speed up the research in this area.
\end{abstract}


\section{Introduction}
\label{sec:intro}

Gesture is a remarkable interaction way for Human Computer Interaction (HCI), which is a conventional non-verbal communication method. It is one type of pervasive body language that can be used for communication. However, with the development of the gaming interaction and wearable device, precise finger gesture have more advantages than body gesture, especially for control devices\cite{Hauslschmid2015} The finger movement is one of the micro-gestures that can accurately manipulate the device. Kinect and RBG-D camera are popular sensors for gaming in the Augmented Reality (AR) and Virtual Reality (VR) community with its low-cost been a major advantage, as well as its immersive user experience and usability~\cite{Zhang2012}. {\color{black}{There displays support the 3D gesture systems and need free space to support  flexible interaction~\cite{Wang2017}.}}. However, these systems lack the ability to capture quality and accurate objects which could be seen as one of its major drawbacks\cite{Aouada2013}. Recently, some new research from Leap Motion \cite{Marin2016} and Google Soli Project~\cite{Lien2016} created new techniques for 3D detection that has huge potentials for success. Holoscopic 3D (H3D) imaging system is a novel potential technique which can satisfy the higher demand of user interactive experience. Detection of precision 3D micro-gesture can make use of the wide view coverage the of Holoscopic 3D camera to capture accurate finger movement~\cite{Swash2014}.

This paper aims to use the H3D imaging system to create a unique 3D micro-gesture database, further to promote the gesture recognition. Even using H3D system is depart from precedents in capture and recognition of the gesture area, this technology was mature to use the 3DTV and display area. Meanwhile, the H3D system also supports the RGB high quality dynamic and static data, and it is renowned for high accuracy and true 3D to excellence than traditional 3D capture devices.   
\begin{figure*}[htb]
\begin{minipage}[b]{1.0\linewidth}
    \centering
\includegraphics[width=15cm]{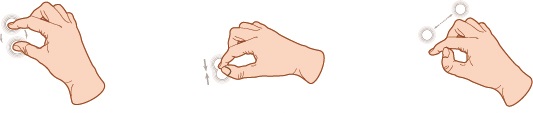}
  \centerline{(a) Button \ \ \ \ \ \ \ \ \ \ \ \ \ \  \ \ \ \ \  \  \ \ \ \   \ \ \ \ \ \ \ \ \ \ \ \ \ \  \ \ \ \ \  \  \ \ \ \  \ \  (b) Dial \ \ \ \ \ \ \ \ \ \ \ \ \ \  \ \ \ \ \  \  \ \ \ \ \ \ \ \ \ \ \ \ \ \ \ \ \ \  \ \ \ \ \  \  \ \ \ \ \ \      (c) Slider }\medskip
  \end{minipage}
\caption{Three different types of finger micro-gestures studied in this database.}
\label{micro-gesture}
\end{figure*}

\section{Related Work}
\label{sec:format}

HCI appeared early in 1983~\cite{Zhang2013}, which use multiple modalities such as voice, gestures (e.g. body, hand, arm, finger, etc.). For example, Siri~\cite{Bellegarda2014} is a very popular voice-based interface. However nature gesture is another way to interact with the computer. The trend of the HCI is user experience of intuitionistic and effective~\cite{Wang2017}. Gesture is a touchless, non-intrusive method for HCI, and it is represented as the diverse type of the gestures~\cite{Pisharady2015}. Manipulative type of gesture appears the most popular one from the preciously literatures. The aim is to control entity being manipulated through the actual movements of the gesturing hand and arm~\cite{Karam2005}. Hand as a direct input device is more and more popular, as one of the outstanding interaction methods.

The Kinect and RGB-D camera are very popular in recent years due to the benefits of Kinect and RGB-D camera that have low cost and wide availability as a sensor~\cite{Ren2012} to capture gestures. However, RGB-D camera suffers from the underside artefacts such as the edge inaccuracies, low object remission and holes owning~\cite{Aouada2013}. The Kinect sensor offers the information of the depth measurement and create coordinates of the 3D objects. Although the abundant development toolkits can support the human body recognition, the weakness is its lacking ability to capture the flexible and robust mechanism to perform high-level gesture~\cite{Ibanez2014}.

Leap Motion (LM)~\cite{Potter2013} is a device that can be used to detect the hand and finger dynamic movements through its API software. The API has the robust pre-processing function which can reduce the complexity of the user control. However, LM is a monocular video sensor which is a challenging for capturing the abundant dynamic hand gestures and finger micro movements~\cite{Lu2016}. 

Holoscopic 3D camera is a single aperture sensor not only to represent the real-time and represents a true volume spatial optical model of the object scene but also to record the viewing natural continuous parallax 3D objects within a wide viewing zone~\cite{Zhao2013}. It provides a new way to capture micro-gestures.

Isaac et al.~\cite{Wang2017} presents a review summary of  21  gesture datasets from previous research and publicly datasets, in which 7 databases are for hands. Most datasets are recorded using to the Kinect or RGB-D camera as the sensor.

In order to the support the diversification of the gesture recognition and encourage development of the human computer interaction, We propose a new 3D gesture database included the three ubiquitous micro gestures that are most the popular ones used in the Google Soli project. Those are intuitive and unobtrusive manipulative gesture. This database is not only include the continuous dynamic data, but also contained the abundant static data to support the 3D micro-gesture recognition. 

\section{Database Construction}

\subsection{Micro-gestures}
There are many micro-gestures that can be used for control in AR and VR applications. In this research, three intuitive micro-gestures are selected refer to the Google Soli project~\cite{Lien2016} as shown in Fig.1. The three gestures are based on the human intuitiveness when they try to control display. For instance, the button gesture is the submission, dial gesture shows that user want to slight change the current situation, and the slider gesture is to express the slide up or down to adjust the volume and options. This three gesture belongs to the manipulative type of the gesture, which use to touch- less control the devices or simulation console. 

\subsection{H3D imaging technology}
H3D imaging technology is success for use in the 3D cinema, 3D-capable televisions and broadcasters. The H3D camera used here is built from the 3D Vivant Project (3D Live Immerse Video-Audio Interactive Multimedia)~\cite{Aggoun2013} and the purpose is to capture high quality 3D images. The developed camera includes micro-lens array, relay lens and digital camera sensors. The principle of the holoscopic 3D imaging is shown in Fig.2. The 3D holoscopic image's spatial sampling is determined by the number of lens. It shows that the captured 2D lenslet array views is slight different angle than its neighbor and reconstructed image in relay~\cite{Aggoun2013}. The detailed parameters of the camera are shown in Fig.3. %The full parallax colour 3D imaging system have big potential for gesture recognition. 

\begin{figure}[htb]
    \centering
\includegraphics[width=7.5cm]{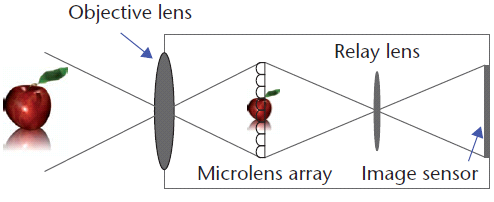} 
\caption{Principle of the holoscopic 3D camera. The microlens array is placed between objective and relay lens to produce fly eye style images.~\cite{Aggoun2013}}
\label{(a)Capture.(b)Replay}
\end{figure}
%The HoMG database includes the 3 micro gesture movements. And this database have revolutionary features.
%1) High quality Multi-data. 
Holoscopic 3D camera sensor has unique optical components which support the continuous parallax RGB image system, contain the depth information viewpoint images. The figure show the H3D imaging having the full color with full parallax. H3D imaging is comprised of the 2D array of micro images. 

The H3D sensor is a crucial requirement for the capture of the objects. This database uses the H3D imaging system to support the dynamic and static RGB data. And it’s not only can record the continuous motion, but repetitive lens array can extract different angles viewpoint images. The uniqueness to encourage the innovation of gesture capture and recognition.

%2) Comprehensive consideration. In order to all-sided data, we invited the over 40 people to join this shoot. And no race and age limited. For shoot part, we present the two distances, two backgrounds, and recorded the right and left hand.   
%3) Non-intrusive micro gestures. This database proposed three intuitive gestures refer to the Google soli project. 
\begin{figure}[htb]
\centering
\includegraphics[width=8.5cm]{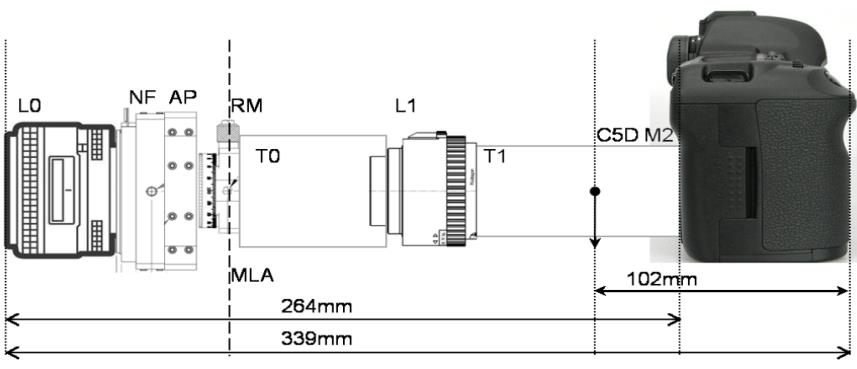}
\caption{Assembled holoscopic 3D camera.}
\label{Assembled holoscopic 3D camera}
\end{figure}

\subsection{Recording Setup}
\label{sec:recording}

\begin{figure}[htb]
\centering
\includegraphics[width=8.5cm]{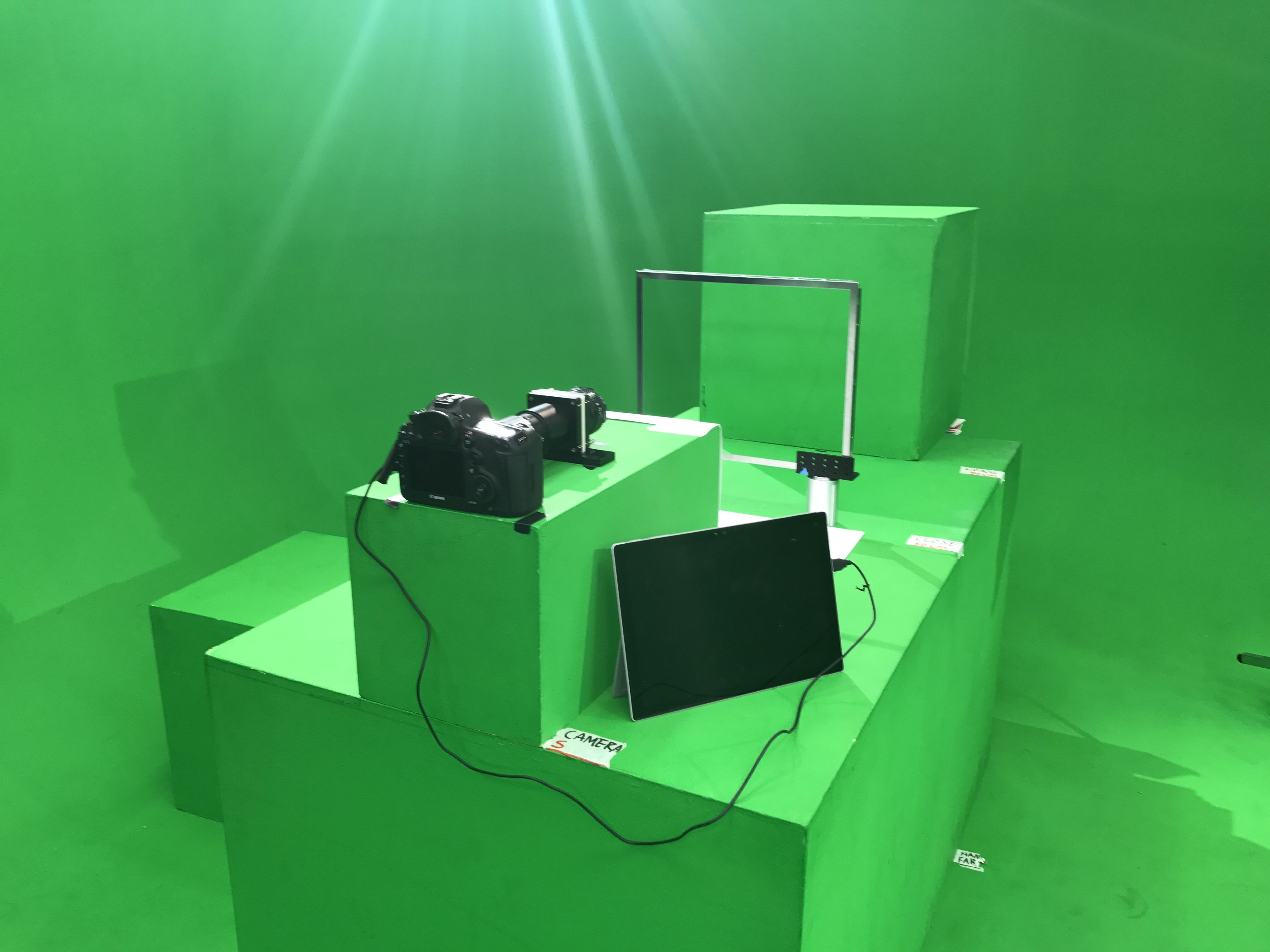}
\caption{Data acquisition setup.}
\label{Experimental Setup}
\end{figure}

The recording H3D gesture place, as in Fig.4. A green screen room is used for the recording where it can offer clear and professional recording background to reduce noise. Before the recording, the holoscopic 3D camera adapter and surface are set up in advance. Canon 5D camera is used and the film speed is ISO200, shutter 1/250. Holoscopic 3D camera adapter is calibrated and the lens are corrected. 

Considering to the influence from distances, angles, and backgrounds, we prepared 4 positions for participants.  Two positions are the close and far locations where the objective lens set to 45cm and 95cm. The other two positions are from left and right hand side for the convenience of the participants. In the close position, we set a hollow frame to help the participant to find the 3D micro-gesture capture zone. 
%Before recording the movement, we predesignate the three gestures: Button, Dial, and Slider. 
We remind the participants the gesture name while record their finger movements. The participants can perform their micro-gesture at their own speed. The recording are done during the time around 15 minutes for one participant.

We prepared two different colour backgrounds, two different distances of close and far from end of the camera lens to gesture area. The recorded imaging resolution is 1086x1902, and the micro lens is 28x28. Participants are successively stand each pre-established position to play three gestures around 3-5 seconds. The three gestures are involved button, dial and slider.

\begin{table}[h]
\centering
\caption{Detailed information about the data acquisition.}
\label{my-label}
\vspace{0.2cm}
\begin{tabular}{|l|l|}
\hline
Parameters       & Detailed information   \\
\hline
Micro-gesture       &    Button (B), Dial (D), Slider (S) \\ \hline
Participants & Male (33), Female (17)      \\ \hline
Hand               & Right (R), Left (L)                                                  \\ \hline
Distance & Close ( 45cm), Far ( 95cm)       \\ \hline
Background & Green (G), White (W)      \\ \hline
Camera & Canon 5D \\ \hline
Image resolution &  1902 x 1086      \\ \hline
Lens array & 28 x 28      \\ \hline
Shutter speed & 1/250 \\ \hline
Film speed & ISO200 \\ \hline
Frame rate &  25      \\ \hline
Recording length &  Between 2 and 20 Sec.      \\ \hline
\end{tabular}
\end{table}

\subsection{Participants}

In total, 40 participants attended the recordings including 17 female participants and 33 male participants, who all read the participant information sheet guidance and sign the research ethics application forms before the recording. There is no any limited of age and race for the participants and We respect the participants’ will. Some participants wear married rings and watches during the recording on finger movements. These increase the data’s noise and bring more challenges. We recorded the participants’ double hands in order to increase the diversity of the data. The detailed information about the data acquisition is summarized in Table 1.

\subsection{HoMG Database}
For the data collection, the recordings from 40 participants are selected to make the HoMG database. The recordings were done under different conditions. One participants were recorded 24 videos. In total, 960 videos are included in the database. 

For micro-gesture recognition, it can be done based on single image or can be done from a short video. So this database was divided into two subsets: image based and video based micro-gesture subsets.

\subsubsection{Video subset}
There are 40 subjects and each subject has 24 videos due to the different setting and three gestures. For each video, the frame rate is 25 frames per second and length of videos are from few seconds to 20 seconds and not equally. The whole dataset was divided into 3 parts. 20 subjects for training set, 10 subjects for development set and another 10 subjects for testing set. In this way, the micro-gesture recognition are person independent.

\subsubsection{Image subset}
Video can capture the motion information of the micro-gesture and it is a good way for micro-gesture recognition. However, it needs more data and take long time. It is very interesting to see whether it is possible to recognise the micro-gesture from a single image with high accuracy.  
 
From each video recording, different number of frames were selected as the still micro-gesture images. In total, there are 30635 images selected. The whole dataset were split into three partitions: a Training, Development, and Testing partition. There are 15237 images in the training subsets of 20 participants with 8364 in close distance and 6853 in far distance. There are 6956 images in the development subsets of 10 participants with 3077 in close distance and 3879 in far distance. There are 8442 images in the testing subsets of 10 participants with 3930 in close distance and 4512 in far distance. 

%totally collected the 720 videos and over 144000 frames which separate to the 24 labels. We abbreviated each video’s label and implication of the label detailed is like below table.

\section{Initial investigation on Micro-Gesture Recognition}
\label{sec:recognition}

The initial investigation are carried out independently for video and image based micro-gesture recognition study. We would like to see how high performance can be achieved from each way. 

%divided into two parts, one is the dynamic result of the dataset classify by extracting the dynamic features of the data; the other part is the static result of the dataset classify the static feature of the dataset. The dynamic feature is extracted directly from videos of gestures and the static feature is extracted from selected images in each video.

%There are 30 subjects in dataset, and each of them is asked to perform 24 videos of gestures including 3 different gestures, button, dial and slider. The subjects are asked to perform the 3 gestures in two different solid distances, one is far from camera and another is close; in two different colour of background, white and green; and with both hands of subjects. The schematic of different videos for subjects are shown in figure 1 {???}.

% \begin{figure}[htb]
%   \centering
%   %\includegraphics[scale=1.0]{tu2.JPG}
%   \caption{Flowchart of experimental system}
%   \label{figurelabel}
% \end{figure}

\subsection{Video based micro-gesture recognition}
There are many good features that can be extracted from each video to capture the movement of the fingers. Here LBPTOP \cite{Zhao:2007:DTR:1263144.1263469} and LPQTOP \cite{6504484} are selected. These features can not  only calculate the distribution of the local information of each frame, but also the distribution of finger movements along to the time. From each video, the frame size was reduced to 74x42 from 1920x1080 firstly, then a feature vector with dimension of 768 is extracted using LBPTOP and LPQTOP for the classification. For the classification, it is a three class classification problem. There are lots of classifiers available. Here, most popular ones such as k-NN, Support Vector Machines (SVM) and Naive Bayes classifiers are chosen for comparison purpose.

Table 2 shows the accuracy using three different classifiers under different distance on video based micro-gesture recognition. From this table, it can be seen that LPQTOP is better than LBPTOP for feature extraction. SVM is better than k-NN and Naive Bayes classifiers in most cases. The accuracy on close distance is better than far distance because the detailed information of the finger movement can be captured. For the testing set, both training and development sets were used for training. Overall, 66.7\% accuracy can be achieved even use the feature extraction methods from all videos in the testing set.

\begin{table}
\caption{Recognition accuracy (\%) of video based micro-gesture recognition on development (Dev.) and testing sets using k-NN, SVM and Naive Bayes classifiers.}
\label{tab:image}
\vspace{0.2cm}
\begin{center}
\begin{tabular}{|c|c|c|c|c|c|}
\hline
\multirow{2}{*}{Dataset}&\multirow{2}{*}{Distance} & \multirow{2}{*}{Feature} & \multicolumn{3} {c|}{Classifier}\\ \cline{4-6}
%\hline
& &  & k-NN & SVM & Bayes\\
\hline
\multirow{6}{*}{Dev.}&\multirow{2}{*}{Close} & LBPTOP &53.3  & 68.3&  52.5\\
\cline{3-6}
& & LPQTOP &56.7  &66.7  &63.3 \\
\cline{2-6}
&\multirow{2}{*}{Far} & LBPTOP &40.8  &53.3 & 47.5 \\
\cline{3-6}
& & LPQTOP &50.8  &55.8  &49.2\\
\cline{2-6}
&\multirow{2}{*}{All} & LBPTOP &44.5   &52.9  &47.9 \\
\cline{3-6}
& & LPQTOP &47.9  &60.4  &51.3  \\
\hline
\multirow{6}{*}{Test}&\multirow{2}{*}{Close} & LBPTOP &56.7 &53.3 &40.8 \\
\cline{3-6}
& & LPQTOP &67.5  &73.3  &65.8 \\
\cline{2-6}
&\multirow{2}{*}{Far} & LBPTOP &55  &55 &50.8 \\
\cline{3-6}
& & LPQTOP &51.7 &65.8 & 58.3 \\
\cline{2-6}
&\multirow{2}{*}{All} & LBPTOP & 53.3 &59.5 &45.4 \\
\cline{3-6}
& & LPQTOP &60.4 &\bf{66.7} &57.5 \\
%SVM & 49.3\% & 64.9\%\\
%\hline
%k-NN & 35.8\% & 46.5\%\\
%\hline
%Subspace Discriminant & 51.7\% & 65.6\%\\
\hline
\end{tabular}
\end{center}
\end{table}

%%%
\begin{table}
\caption{Recognition accuracy (\%) of image based micro-gesture recognition on development (Dev.) and testing sets using k-NN, SVM and Naive Bayes classifiers under different distance conditions.}
\label{tab:image}
\vspace{0.2cm}
\begin{center}
\begin{tabular}{|c|c|c|c|c|c|}
\hline
\multirow{2}{*}{Dataset}&\multirow{2}{*}{Distance} & \multirow{2}{*}{Feature} & \multicolumn{3} {c|}{Classifier}\\ \cline{4-6}
%\hline
& &  & k-NN & SVM & Bayes\\
\hline
\multirow{6}{*}{Dev.}&\multirow{2}{*}{Close} & LBP & 40.9 & 44.3 & 46.0\\
\cline{3-6}
& & LPQ & 43.4 & 45.0 & 42.8\\
\cline{2-6}
&\multirow{2}{*}{Far} & LBP & 35.9 & 32.1 &37.4\\
\cline{3-6}
& & LPQ & 36.7 & \bf{52.6} & 47.5\\
\cline{2-6}
&\multirow{2}{*}{All} & LBP & 41.0 & 35.0 & 39.6\\
\cline{3-6}
& & LPQ & 32.9 & 51.6 & 50.6\\
\hline
\multirow{6}{*}{Test}&\multirow{2}{*}{Close} & LBP & 49.7
& 33.6 & 45.4\\
\cline{3-6}
& & LPQ & 44.1 & 46.4 & 39.7\\
\cline{2-6}
&\multirow{2}{*}{Far} & LBP & \bf{50.9} & 37.7& 47.2\\
\cline{3-6}
& & LPQ & 34.6 & 51.6& 50.0\\
\cline{2-6}
&\multirow{2}{*}{All} & LBP & 44.7 & 48.9 & 44.7\\
\cline{3-6}
& & LPQ &46.8 & 50.9 & 46.8\\
%SVM & 49.3\% & 64.9\%\\
%\hline
%k-NN & 35.8\% & 46.5\%\\
%\hline
%Subspace Discriminant & 51.7\% & 65.6\%\\
\hline
\end{tabular}
\end{center}
\end{table}

\subsection{Image based micro-gesture recognition}
 
%As for Close setup; 8364 images is selected for Train, 3077 images for Development, and 3930 images for Testing. As for Far setup, 6853 images is selected for Train, 3879 images for Development and 4512 images for Test. As for concatenation between Close and Far setup, 16153 images selected for Train, 8046 images for Development and 6415 for Testing. 

For each image, 2D texture features such LBP \cite{Ojala1996} and LPQ \cite{Ojansivu2008} were extracted to represent each image. These two features captured the edge and local information of the 2D image in different ways and form a histogram feature vector with dimension of 256. Popular classification methods such as k-NN, SVM and Naive Bayes classifiers were used for recognising the three different micro-gestures.  

Table 3 should the experimental results on video based micro-gesture recognition by training on the training set and tested on development and testing subsets. From this table, it can be seen that for most of the classifications, around 50\% accuracy can be achieved.

\section{CONCLUSIONS AND FUTURE WORKS}

\subsection{Conclusions}
This paper introduces a unique holoscopic 3D micro-gesture database (HoMG), which is recorded under different settings and conditions from 40 participants. The data recording uses the similar the H3D system of fly viewing to capture the participants’ precise finger movements. The H3D imaging system supports robust 3D depth micro lens array to capture dynamic and static information. The HoMG database has 3 unobtrusive manipulative gestures in two different backgrounds, two different distances, left and right hands. These micro-gestures can be used to control multifarious displays. This database would speed up the research in this area. 

The database is further divided into video and image subsets. Initial investigation on micro-gesture recognition is conducted. For the comparison, video based method achieved better performance as it has dynamic finger movement information in the data. However, this method needs much more data and computing time. Image based method is convenient for the user and might have more applications, especially on the portable devices. Even with the standard 2D feature extraction methods and basic classification methods, 72.5\% recognition accuracy can be achieved for micro-gesture videos and over 50\% accuracy for micro-gesture images. This baseline methods and results will give a foundation for other researchers to explore their methods. 

\subsection{Future works}
From the initial investigation, it can be seen that the recognition accuracy can reach around 66\% even just using the 2D image processing methods. For 3D image processing methods, such as extracting the different viewing point images and extract 3D information of the micro-gesture, high accuracy will be achieved. This will be our future works. 

\section{ACKNOWLEDGEMENTS}
We gratefully acknowledge the support of NVIDIA Corporation with the donation of the Tesla K40 GPU used for this research.

%\section{LAST PAGE}
%\label{sec:print}

%If the last page of your paper is only partially filled, arrange the columns so that they are evenly balanced (if possible), rather than having
%one long column, as illustrated on this page in the template.pdf output.

%To start (in \LaTeX) a new column (but not a new page) and help balance the
%last-page column lengths, you can use the sequence of commands 
%\verb|\vfill\pagebreak|, as shown in the template.tex file. 
%You must place this pair of commands just after the text in the first column where you think you must stop this column. You may have to try a few positions to get the best balance.

%\vfill
%\pagebreak

%-------------------------------------------------------------------------
%\bibliographystyle{plain}
%\bibliography{sample-bibliography}

\begin{thebibliography}{10}

\bibitem{Aggoun2013}
Amar Aggoun, Emmanuel Tsekleves, Mohammad~Rafiq Swash, Dimitrios Zarpalas,
  Anastasios Dimou, Petros Daras, Paulo Nunes, and Lu{\'{i}}s~Ducla Soares.
\newblock {Immersive 3D holoscopic video system}.
\newblock {\em IEEE Multimedia}, 20(1):28--37, 2013.

\bibitem{Aouada2013}
Djamila Aouada, Bj{\"{o}}rn Ottersten, Bruno Mirbach, Frederic Garcia, and
  Thomas Solignac.
\newblock {Real-time depth enhancement by fusion for RGB-D cameras}.
\newblock {\em IET Computer Vision}, 7(March):335--345, 2013.

\bibitem{Bellegarda2014}
Jerome~R. Bellegarda.
\newblock {\em Spoken Language Understanding for Natural Interaction: The Siri
  Experience}, pages 3--14.
\newblock Springer New York, New York, NY, 2014.

\bibitem{Hauslschmid2015}
Renate H{\"{a}}uslschmid, Benjamin Menrad, and Andreas Butz.
\newblock {Freehand vs . Micro Gestures in the Car : Driving Performance and
  User Experience}.
\newblock 0336:159--160, 2015.

\bibitem{Ibanez2014}
Rodrigo Iba{\~{n}}ez, {\'{A}}lvaro Soria, Alfredo Teyseyre, and Marcelo Campo.
\newblock {Easy gesture recognition for Kinect}.
\newblock {\em Advances in Engineering Software}, 76:171--180, 2014.

\bibitem{6504484}
B.~Jiang, M.~Valstar, B.~Martinez, and M.~Pantic.
\newblock A dynamic appearance descriptor approach to facial actions temporal
  modeling.
\newblock {\em IEEE Transactions on Cybernetics}, 44(2):161--174, Feb 2014.

\bibitem{Karam2005}
Maria Karam and m.~c. Schraefel.
\newblock {A Taxonomy of Gestures in Human Computer Interactions}.
\newblock {\em Technical Report, Eletronics and Computer Science.}, pages
  1--45, 2005.

\bibitem{Lien2016}
Jaime Lien, Nicholas Gillian, M~Emre Karagozler, Patrick Amihood, Carsten
  Schwesig, Erik Olson, Hakim Raja, Ivan Poupyrev, and Google Atap.
\newblock {Soli: Ubiquitous Gesture Sensing with Millimeter Wave Radar}.
\newblock {\em ACM Trans. Graph. Article}, 35(10), 2016.

\bibitem{Lu2016}
Wei Lu, Zheng Tong, and Jinghui Chu.
\newblock {Motion Controller}.
\newblock 23(9):1188--1192, 2016.

\bibitem{Marin2016}
Giulio Marin, Fabio Dominio, and Pietro Zanuttigh.
\newblock {Hand gesture recognition with jointly calibrated Leap Motion and
  depth sensor}.
\newblock {\em Multimedia Tools and Applications}, pages 14991--15015, 2016.

\bibitem{Ojala1996}
Timo Ojala, Matti Pietik{\"{a}}inen, and David Harwood.
\newblock {A comparative study of texture measures with classification based on
  featured distributions}.
\newblock {\em Pattern Recognition}, 29(1):51--59, 1996.

\bibitem{Ojansivu2008}
Ville Ojansivu and Janne Heikkil{\"{a}}.
\newblock {\em {Blur Insensitive Texture Classification Using Local Phase
  Quantization}}, pages 236--243.
\newblock Springer Berlin Heidelberg, Berlin, Heidelberg, 2008.

\bibitem{Pisharady2015}
Pramod~Kumar Pisharady and Martin Saerbeck.
\newblock {Recent methods and databases in vision-based hand gesture
  recognition: A review}.
\newblock {\em Computer Vision and Image Understanding}, 141:152--165, 2015.

\bibitem{Potter2013}
Leigh~Ellen Potter and Jake Araullo.
\newblock {The Leap Motion controller : A view on sign language}.
\newblock pages 175--178, 2013.

\bibitem{Ren2012}
Gang Ren and Eamonn O'Neill.
\newblock {3D Marking menu selection with freehand gestures}.
\newblock In {\em IEEE Symposium on 3D User Interfaces 2012, 3DUI 2012 -
  Proceedings}, pages 61--68, 2012.

\bibitem{Swash2014}
M.~R. Swash, O.~Abdulfatah, E.~Alazawi, T.~Kalganova, and J.~Cosmas.
\newblock {Adopting multiview pixel mapping for enhancing quality of holoscopic
  3D scene in parallax barriers based holoscopic 3D displays}.
\newblock {\em IEEE International Symposium on Broadband Multimedia Systems and
  Broadcasting, BMSB}, pages 1--4, 2014.
\vfill
\pagebreak
\bibitem{Wang2017}
Isaac Wang, Mohtadi~Ben Fraj, Pradyumna Narayana, Dhruva Patil, Gururaj Mulay,
  Rahul Bangar, J~Ross Beveridge, Bruce~A Draper, and Jaime Ruiz.
\newblock {EGGNOG : A continuous , multi-modal data set of naturally occurring
  gestures with ground truth labels}.
\newblock pages 414--421, 2017.

\bibitem{Zhang2013}
Chenyang Zhang, Xiaodong Yang, and Yingli Tian.
\newblock {Histogram of 3D Facets: A characteristic descriptor for hand gesture
  recognition}.
\newblock In {\em 2013 10th IEEE International Conference and Workshops on
  Automatic Face and Gesture Recognition, FG 2013}, 2013.


\bibitem{Zhang2012}
Zhengyou Zhang.
\newblock {Microsoft kinect sensor and its effect}.
\newblock {\em IEEE Multimedia}, 19(2):4--10, 2012.

\bibitem{Zhao:2007:DTR:1263144.1263469}
Guoying Zhao and Matti Pietikainen.
\newblock Dynamic texture recognition using local binary patterns with an
  application to facial expressions.
\newblock {\em IEEE Trans. Pattern Anal. Mach. Intell.}, 29(6):915--928, June
  2007.

\bibitem{Zhao2013}
Yue Zhao, Yongtao Wei, Xiaoyu Cui, Luxuan Qu, Lin Liu, Yusong Wang, Jingang
  Wang, Xiao Xiao, Hong Hua, Bahram Javidi, M.~R. Swash, A.~Aggoun,
  O.~Abdulfatah, B.~Li, J.~C. Fernandez, E.~Tsekleves, J~C Fern{\'{a}}ndez,
  E~Alazawi, E.~Tsekleves, Gorkem Saygili, C.Goktug Gurler, A.~Murat Tekalp,
  Kunio Sakamoto, Rieko Kimura, Miwa Takaki, Martin Reˇ, Guo~Jiao Lv, Wu~Xiang
  Zhao, Da~Hai Li, Qiong~Hua Wang, Sung~Kyu Kim, Dong~Wook Kim, Min~Chul Park,
  Jung~Young Son, Minju Kim, B.~Kaufmann, M.~Akli, Hsin-jung Chen, Feng-hsiang
  Lo, Fu-chiang Jan, and Sheng-dong Wu.
\newblock {3D images compression for multi-view auto-stereoscopic displays}.
\newblock {\em IEEE/OSA Journal of Display Technology}, 3(11):1165--1168, 2013.

\end{thebibliography}

\end{document}